\documentclass[%
 reprint,
superscriptaddress,
nofootinbib,
amsmath,amssymb, aps,
]{revtex4-1}

\usepackage{graphicx}
\usepackage{dcolumn}
\usepackage{bm}
\usepackage{hyperref}
\usepackage{adjustbox}
\usepackage{footmisc}

\begin{document}

\preprint{APS/123-nuclear structure}

\title{A way forward towards improvement of tensor force in \textit{pf}-shell}
\author{Kanhaiya Jha}
\email{kanhaiya.jha@iitrpr.ac.in}
\affiliation{Department of Physics, Indian Institute of Technology Ropar, Rupnagar, 140001, Punjab, India}
\author{Pawan Kumar}
\email{pawan.kumar@iitrpr.ac.in}
\affiliation{Department of Physics, Indian Institute of Technology Ropar, Rupnagar, 140001, Punjab, India}
\affiliation{Department of Physics, Aksaray University, 68100 Aksaray, Turkey}
\author{Shahariar Sarkar}
\affiliation{Department of Physics, Indian Institute of Technology Ropar, Rupnagar, 140001, Punjab, India}
\author{ P. K. Raina  }
\affiliation{Department of Physics, Indian Institute of Technology Ropar, Rupnagar, 140001, Punjab, India}
\author{Sezgin Aydin}
\affiliation{Department of Natural and Mathematical Sciences,
Engineer Faculty, Tarsus University, 33400, Tarsus, Turkey}

\begin{abstract}
In many shell model interactions, the tensor force monopole matrix elements are observed to retain systematic trends originating in the bare tensor force. In this work, however, we note for GX-interactions of $pf-$shell that the seven out of ten T = 1 tensor force monopole matrix elements  do not share these systematic. We ameliorate this disparity making use of Yukawa-type tensor force and spin-tensor decomposition. Furthermore, we modify the single-particle energy of $1p_{3/2}$  orbit and two TBMEs of $0f$-orbit, and test the revised interaction from Ca to Ge isotopes with various physics viewpoints. The results are found to be satisfactory with respect to experimental data.
\end{abstract}

\maketitle


\section{Introduction}
The nuclear shell-model is one of the most successful theoretical frameworks which has been extensively used for understanding the nuclear structure properties. In its long-run success, the employed effective interactions have played a pivotal role \cite{Caurier,Otsuka1,Brown1}. The effective interactions are generally derived by modifying the bare nucleon-nucleon (NN) interactions using the microscopic and phenomenological approaches in order to incorporate the short-range correlations, in-medium, and many-body force effects \cite{Kuo,Jensen,Dean,Chung,Poves1}. The monopole matrix elements
\begin{equation}
\label{E1}
\bar{V}^{T}_{jj'} = \frac{\sum_{J} (2J+1) \langle jj'|V| jj'\rangle_{JT} }{\sum_{J} 2J+1} 
\end{equation}
are the key ingredients of the effective interaction \cite{Poves1, Zuker, Dufour}, and small modification in them notably improves the description of experimental data. For example, effective interactions - KB3G \cite{Poves2}, and GXPF1B \cite{Honma1} are derived from the monopole modification to give a good description of single-particle and collective properties of $pf-$shell nuclei. In the present study, we work in $pf-$shell and are mainly paying attention to the modified form of GXPF1B interaction, which hereafter is denoted by GX1B1. The $\bar{V}^{T=1}_{p3/2,f5/2}$ matrix elements of these two GX-interactions differ by -0.15 MeV that was suggested to correctly reproduce the E$(2^{+}_{1})$ of $^{54}$Ca \cite{Steppenbeck}.  

In recent years, the different components of NN interaction, i.e, central, spin-orbit and tensor force, have gained a lot of interest to understand the cause of shell evolution at unbalanced ratio of proton-to-neutron in exotic nuclei \cite{Umeya1, Smirnova1, Pawan1, Otsuka2, Utsuno1, Otsuka3}. The bare tensor force originated from $\pi+\rho$ meson exchange, in particular, has attracted everyone interest due to its unique properties \cite{Otsuka2}. It has been demonstrated by the Ostuka and his collaborators that the tensor force majorly cause shell evolution in the whole segr$\grave{e}$ chart \cite{Otsuka2, Utsuno1, Otsuka3}. The bare tensor force monopole matrix elements $\bar{V}^{T}_{jj'} (\mathcal{T})$ have following systematic properties \cite{Otsuka2}: they are attractive for $j_{>} j'_{<}$ ($j_{<} j'_{>}$)\footnote{$j_{>}= l+\frac{1}{2}$ and $j_{<}= l-\frac{1}{2}$ represent spin-up  and spin-down orbits, respectively.} configurations, while repulsive for $j_{>} j'_{>}$ ($j_{<} j'_{<}$) configurations.  Beside this, bare $\bar{V}^{T}_{jj'} (\mathcal{T})$ matrix elements are noted for barely changing against the microscopic renormalization procedures and persisting their systematic properties \cite{Tsunoda}. The numerical analysis based on the spin-tensor decomposition \cite{Kirson,Brown2} shows that $\bar{V}^{T}_{jj'} (\mathcal{T})$ matrix element of well established effective shell-model interaction USDB \cite{Brown3} has the same systematic properties as for the bare tensor force \cite{Wang}. In the present study, we have done the similar analysis  for the $\bar{V}^{T}_{jj'} (\mathcal{T})$ matrix elements of GX1B1 interaction (see Sec. II), and found that seven out of ten T = 1 $\bar{V}^{T = 1}_{jj'} (\mathcal{T})$ matrix elements have irregularities in their properties. For instance, $\bar{V}^{T = 1}_{f7f7} (\mathcal{T})$ is attractive while it was expected to be repulsive. The same tensor force irregularities have also been noted \cite{Wang} for GXPF1A interaction \cite{Honma2}, which is another member of GX-family. In both the effective interactions, this peculiar character may originate from the imprecise normalization of the contribution of higher order in-medium terms and many-body force to their parent interaction - GXPF1 \cite{Honma3} because their renormalized G-matrix interaction owns tensor force properties \cite{Wang}.

In the present study, we have corrected the tensor force irregularities of GX-interactions. We have modified  all ninety-four T = 1 two-body matrix elements of GX1B1. Modification of the interaction on this large scale is a difficult task with full of complications, therefore, there'll always be ambiguity on its practical use. Hence, the revised interaction has been employed for many nuclei with various physics viewpoints. 

This paper is organized as follows. In Sec.~\ref{S2}, we have discussed  (A) spin-tensor decomposition, (B) monopole matrix elements of GX-interactions, and (C) method to bring the tensor force properties in GX-interactions. Results and discussion are presented in Sec.~\ref{S3}. The summary of this work is given in Sec.~\ref{S4}.

\section{Theoretical framework}
\label{S2}
\subsection{Spin-tensor decomposition}
Spin-tensor decomposition is a unique method to break down an effective interaction into its central, spin-orbit and tensor force structure \cite{Kirson,Brown2}. It has been widely used to examine the role of these different forces in level structure \cite{Wang,Umeya2,Klingenbeck,Yoro} and shell evolution \cite{Umeya1, Umeya1, Smirnova1, Pawan1}. In this study, we extend the application of  spin-tensor decomposition by using it as an apropos tool to correct the tensor force disparity of GX-interactions.

In spin-tensor decomposition, the interaction between two-nucleon is defined as the linear sum of the scalar product of configuration space operator $Q$ and spin space operator $S$ of rank $k$ \cite{Kirson}
\begin{equation}
\label{E2}
V = \sum_{k = 0}^{2} V(k) = \sum_{k = 0}^{2} Q^{k}.S^{k},
\end{equation} where, rank k = 0, 1,  and 2 represent central, spin-orbit, and tensor force, respectively. Using the $LS$-coupled two-nucleon wave functions, the matrix element for each $V(k)$ can be calculated from matrix element $V$ \cite{Elhott}
\begin{equation}
\label{E3}
\begin{split}
\langle (ab),LS;JM|V(k)|(cd),L'S';JM \rangle = (2k+1) (-1)^{J} \\
\left \{
  \begin{tabular}{ccc}
  L & S & J \\
  S$'$ & L$'$ & k \\
  \end{tabular}
\right \}
\sum_{J'} (-1)^{J'} (2J'+1)
\left \{
  \begin{tabular}{ccc}
  L & S & J$'$ \\
  S$'$ & L$'$ & k \\
  \end{tabular}
\right \}
\\ \langle (ab),LS;J'M|V|(cd),L'S';J'M \rangle,
\end{split}
\end{equation} where $a = (n_{a}l_{a}) $ is shorthand notation for spherical quantum numbers. In shell model, NN interaction is defined in jj wave functions, therefore, the above expression is needed to be expand in $jj$-basis. This can be done using $9j$-symbol relation between $LS$- and $jj$-coupled wave functions. The final expression can be expressed as \cite{Pawan2}
\begin{multline}
\label{E4}
\langle (j_{a}j_{b});JT |V(k)| (j_{c}j_{d});JT \rangle = 
\dfrac{1}{\sqrt{(1+\delta_{j_{a}j_{b}})(1+\delta_{j_{c}j_{d}})}} \\                          
	\sum_{LSL'S'} (-1)^{J}  (2k+1)
   	\left[
   	\begin{array}{ccc} 
   	l_{a} & 1/2 & j_{a} \\   	l_{b} & 1/2 & j_{b} \\  	 L & S & J
  	 \end{array}
  	\right]        
	\left[
	\begin{array}{ccc} 
 	l_{c} & 1/2 & j_{c} \\  	l_{d} & 1/2 & j_{d} \\  	L' & S' & J
 	\end{array}
 	\right] \\
 	\left\lbrace
 	\begin{array}{ccc}
 	L  & S  & J \\ 	S' & L' & k 
 	\end{array}
 	\right\rbrace  
 	\sum_{J'} (-1)^{J'}  (2J'+1)	
    \left\lbrace
 	\begin{array}{ccc}
 	L  & S  & J' \\ 	S' & L' & k 
 	\end{array}
 	\right\rbrace \\ 
 	  \sum_{j_{a'}j_{b'}j_{c'}j_{d'}}   	 
 	 \left[
 	\begin{array}{ccc} 
 	l_{a} & 1/2 & j_{a'} \\ 	l_{b} & 1/2 & j_{b'} \\ 	L & S & J'
 	\end{array}
 	\right] 		
	\left[
 	\begin{array}{ccc} 
 	l_{c} & 1/2 & j_{c'} \\ 	l_{d} & 1/2 & j_{d'} \\ 	L' & S' & J'
 	\end{array}
 	\right]  \\
 	\sqrt{(1+\delta_{j_{a'}j_{b'}})(1+\delta_{j_{c'}j_{d'}})}  \langle (j_{a'}j_{b'});J'T |V| (j_{c'}j_{d'});J'T \rangle.
\end{multline}

\begin{figure*}[t!]
\includegraphics [height = 9cm, width=25.5cm, trim={1.7cm 3.5cm 0cm 0cm}]{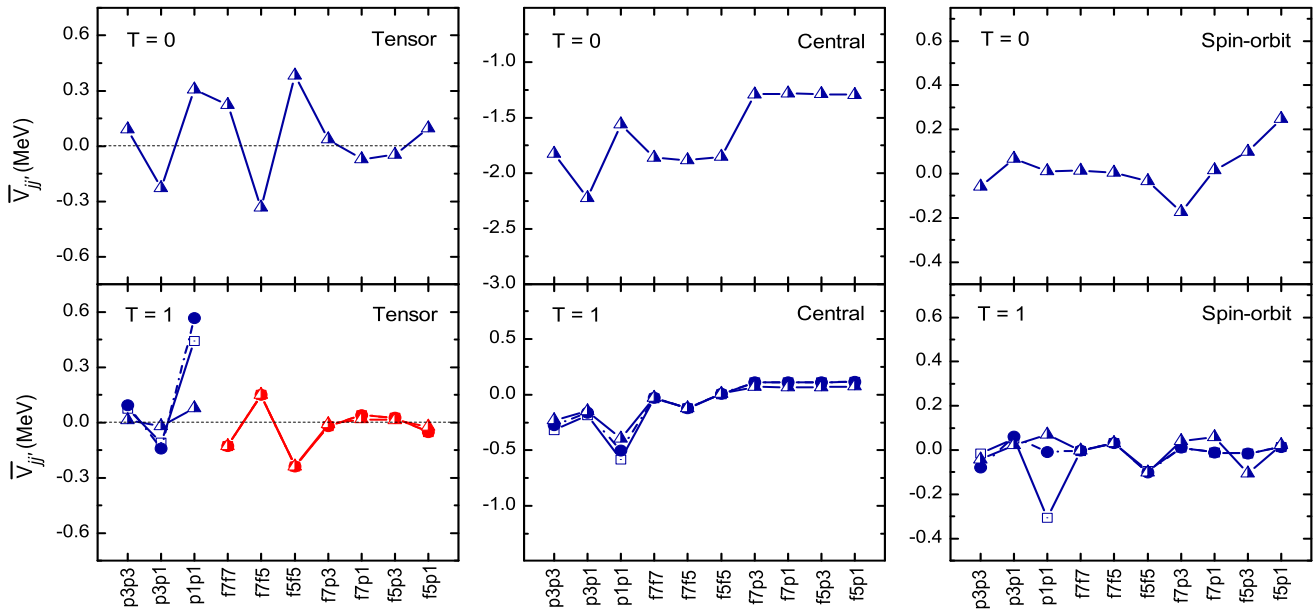}
\caption {Tensor, central, and spin-orbit force monopole matrix elements of GXPF1 (open square), GXPF1A (solid circle), and GX1B1  (half-solid triangle) interactions. Isospin T = 0 matrix elements of these interactions are same, therefore, results of one interaction are shown. Lines are drawn to guide eyes.}
\label{F1}
\end{figure*}

\subsection{Central, spin-orbit, and tensor force monopole matrix elements}

\begin{figure*}
\centering
\includegraphics [height=9cm, width=20cm, trim={-2cm 5.7cm 0cm 0cm}]{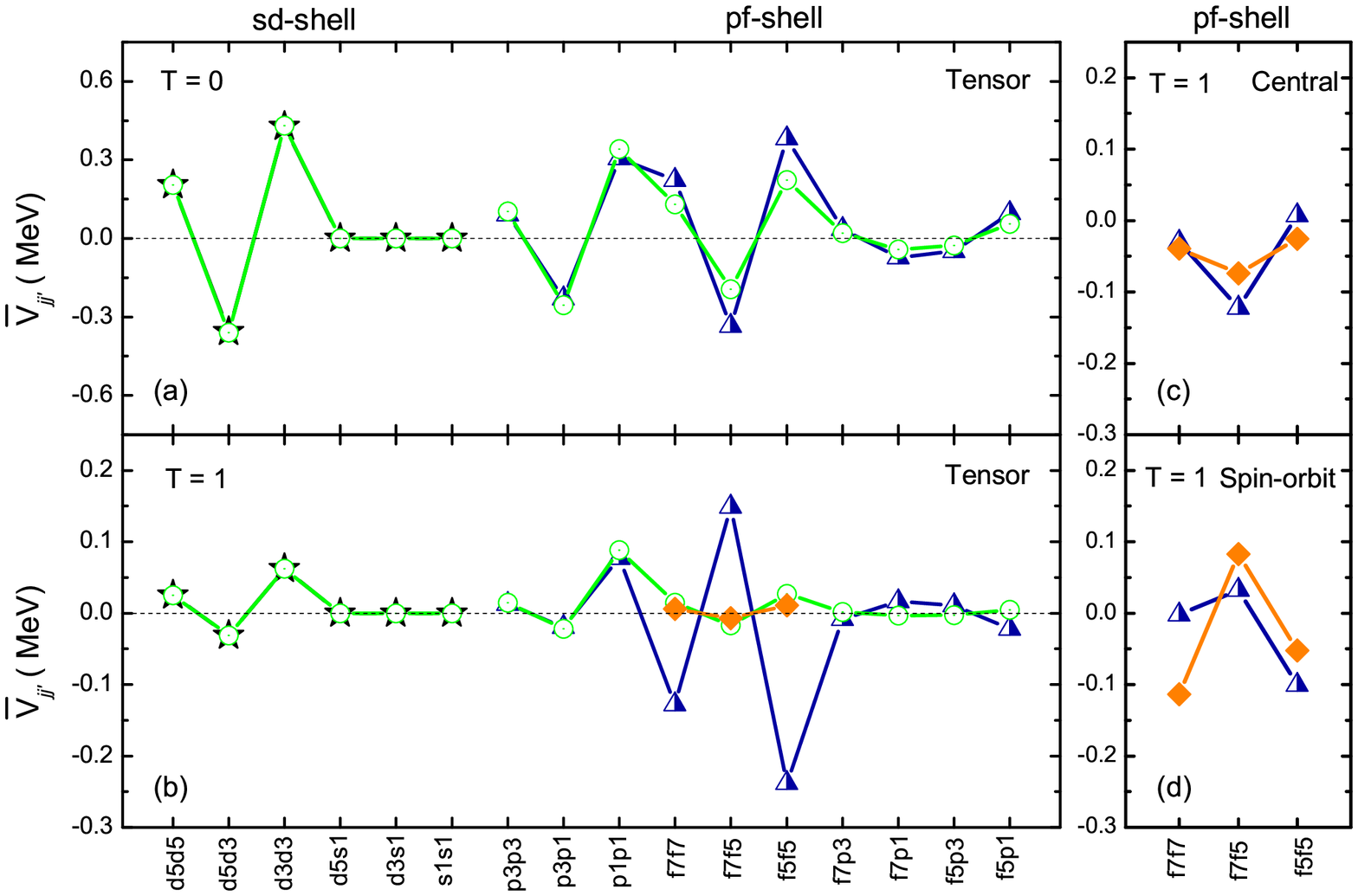}
\caption {Left (a,b): Calculated (open circle) tensor force monopole matrix elements along with those of USDB (solid star) and GX1B1 (half-filled triangle) interactions. Right: Isospin T = 1 (c) central and (d) spin-orbit force monopole matrix elements of GX1B1  interaction. On both sides, solid diamond symbols are used for two \textit{0f}-TBMEs affected monopole matrix elements (see text for details). Lines are drawn to guide eyes.}
\label{F2}
\end{figure*} 

The tensor force monopole matrix elements $\bar{V}^{T}_{jj'} (\zeta)$ of GX1B1 are extracted for both isospin channels using the spin-tensor decomposition. Results are shown on the left side of Fig.~\ref{F1}. In T = 0 channel, all $\bar{V}(\zeta)$ matrix elements have their systematic properties. However, in T = 1 channel, only $\bar{V}_{p3p3}$, $\bar{V}_{p3p1}$, and $\bar{V}_{p1p1}$ have them. The T = 1 $\bar{V}_{f_{7},f_{7}}$, $\bar{V}_{f_{5},f_{5}}$, $\bar{V}_{f_{7},p_{3}}$,  $\bar{V}_{f_{5},p_{1}}$ matrix elements are attractive, and $\bar{V}_{f_{7},f_{5}}$, $\bar{V}_{f_{7},p_{1}}$,  $\bar{V}_{f_{5},p_{3}}$ matrix elements are repulsive. The trends of these matrix elements are expected to be opposite. Similar disparity is also noted for T = 1 $\bar{V}_{jj'} (\zeta)$ matrix elements of GXPF1A \cite{Wang}, see Fig.~\ref{F1}. We have examined the $\bar{V}_{jj'} (\zeta)$ matrix elements of their parent interaction-GXPF1 \cite{Honma3} as well. Interestingly, we find the same disparity in it. Results of GXPF1 interaction are also shown in the same figure.

Since these three GX-interactions have improper tensor force, it may be possible that their other components, i.e., central and spin-orbit force, lack the basic features too. Therefore, we have done the investigation for these forces as well. Results for central force are presented in the middle of Fig.~\ref{F1}. The strong-orbit node and weak spin dependence of central force, reported in Ref. \cite{Otsuka3,Utsuno2}, is descried among the matrix elements, except the $pp-$matrix elements. Although the different property of $p-$orbit matrix elements is consistent with Gaussian-type \cite{Otsuka3} and spin-isospin Yukawa-type \cite{Pawan2} central force, this is a compelling problem that need to be investigated separately. In all three interactions, the observed minor difference is a result of consistent two-body matrix elements modification. 

On the right side of Fig.~\ref{F1}, we show the spin-orbit force monopole matrix elements. The difference in matrix elements of three interactions shows the effects of modification. Unlike the central and tensor force, the systematic properties of spin-orbit force are not yet known, and also cannot be determined from Fig.~\ref{F1}. Therefore, a conclusive statement cannot be made for spin-orbit force in the framework of present work.

\subsection{Tensor force rectification}

In order to rectify the tensor force disparity, we have separately calculated ninety-four T = 1 two-body tensor force matrix elements using tensor force \cite{Otsuka3}
\begin{equation}
V_{\mathcal{T}}= V(r)\sqrt{\frac{24 \pi}{5}}[Y^{(2)}.{(\sigma_{1} X  \sigma_{2} )}^{(2)}](\tau_{1} . \tau_{2}), 
\end{equation}
and replaced them with those of GX1B1. The radial dependency in the above expression is treated with the Yukawa potential \cite{Yukawa}  \begin{equation}
V(r) = -V_{0} \frac{e^{-r/a}}{r/a}, 
\end{equation} for simplicity. Here $V_{0}$ is the strength parameter and `\textit{a}' is the Compton scattering length of pion given as 1.41 fm for $m_{\pi}$ = 139.4 MeV.

In our calculations, the strength parameter is obtained from the fit to $\bar{V}^{T}_{jj'} (\zeta)$ matrix elements of USDB \cite{Brown3}. This fit was first done for $\bar{V}^{T=0}_{jj'} (\zeta)$ matrix elements. These results are presented in Fig.~\ref{F2}(a), which are found to be in a good accordance.  The obtained strength parameter was then used to calculate $\bar{V}^{T=0}_{jj'} (\zeta)$ matrix elements in $pf$-shell, and compared with those of GX1B1. This comparison is also shown in Fig.~\ref{F2}(a). It exhibits that both type of matrix elements are very similar. Hence, it supports the replacement of T = 1 two-body tensor force matrix elements of GX1B1 with the calculated ones. The T = 1 fit in \textit{sd}-shell, and the calculated $\bar{V}^{T=1}_{jj'} (\zeta)$ matrix elements in \textit{pf}-shell are shown in Fig.~\ref{F2}(b). The systematic properties of tensor force in \textit{pf}-shell are found for all calculated matrix elements. Furthermore, calculated $\bar{V}_{p3,p3}$, $\bar{V}_{p3,p1}$ and $\bar{V}_{p1,p1}$ matrix elements are found very similar to those of GX1B1, which manifest that $1p-$orbit matrix elements readjusted for GXPF1B \cite{Honma1} improves the tensor force strength in GX-interactions, consistent with higher order in-medium terms and many-body force effects.

In the next step, we have calculated the level structure of odd and even Ca isotopes as all the modified matrix elements directly belong to them. It was observed that the interaction well predicts spin-parity of ground and excited states. However, it predicts somewhat lower excitation energy than experimental data. The observed difference was in the range of $0.2-0.8$ MeV. Since the central and tensor force now have basic features, the difference between theory and experiment could be helpful in the further exploration of characteristics of spin orbit force in future.

To improve theortical results, we have aimed at the $\nu 1p_{3/2}-\nu 0f_{7/2}$, $\nu 1p_{1/2}-\nu 1p_{3/2}$, and $\nu 0f_{5/2}-\nu 1p_{1/2}$ energy gaps related states \cite{Bhoy,Holt}. We have modified the single-particle energy of $1p_{3/2}$ orbit by -0.221 MeV, and the $ V(7777:61)$ and $ V(7575:61)$ ($V(2j_{a}2j_{b}2j_{c}2j_{d};JT)$) matrix elements by -0.280 MeV and 0.399 MeV, respectively. It was captivating to note that these small modifications had improved the level structure in overall. The level structure of Ca isotopes are discussed in Sec.III. 

It is important to mention that both $0f-$orbit matrix elements were adjusted in such a way that it does not affect the tensor force properties of $\bar{V}_{f7,f7}$, $\bar{V}_{f7, f5}$, and $\bar{V}_{f5,f5}$ matrix elements. The change in magnitude appeared in these monopole matrix elements is displayed in Fig.~\ref{F2}(b). The central and spin-orbit force also got changed for the same monopole matrix elements.  They are shown on the right side of Fig.~\ref{F2}. In central force, the strong-orbit node and weak-spin dependence improves among the matrix elements. In spin-orbit force, the matrix elements change by 0.05 to 0.1 MeV in magnitude. The derived interaction, hereafter, is denoted by GX1R. The TBMEs of this interaction can be obtained by contacting the authors.



\section{Results and Discussion}
\label{S3}
\subsection{Effective singe-particle energy}

A basic aspect of the effective shell-model interaction can be obtained from the systematic study of the shell-evolution in the series of isotopes and isotones \cite{Poves1,Utsuno3,Otsuka4}. In shell-model, the shell-evolution is studied with concept of effective single-particle energy that is defined for an orbit $j$ as \cite{Smirnova1,Smirnova2}
\begin{equation}
\epsilon^{'\rho}_{j}(A)= \epsilon^{\rho}_{j}+ \sum_{j'} n_{j'}^{\rho'}  \bar{V}_{jj'}^{\rho \rho'}(A)
\end{equation} where $\rho$ refers to particle type - proton and neutron, $n_{j'}$ is the number of particles in orbit $j'$, and $\bar{V}$ is monopole matrix element. In the above expression, $A$ denotes the mass-dependence of two-body matrix elements (TBMEs). In GX1R, TBMEs for mass-$A$ nuclei are scaled by the mass correction factor $(42/A)^{0.3}$  \cite{Honma3}.

\begin{figure}
\centering
\includegraphics [height= 7cm, width = \columnwidth, trim = {1cm, 5cm, 7cm, 1.cm}]{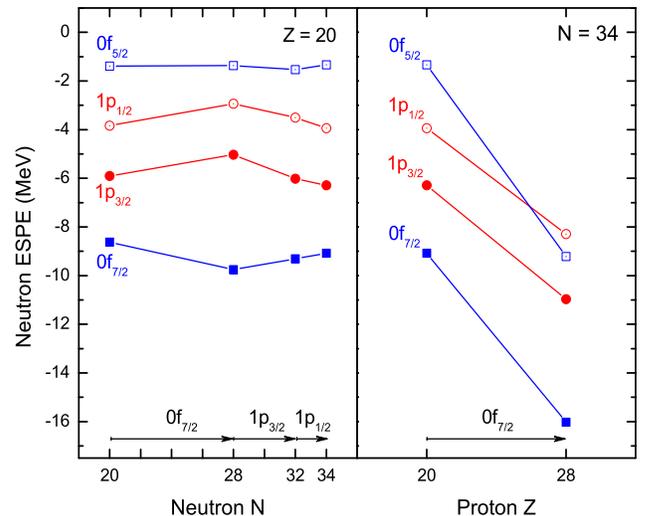}
\caption {Evolution of neutron ESPE of $pf-$orbits in Ca isotopes (left) and N = 34 isotones (right).}
\label{F3}
\end{figure}

In Fig.~\ref{F3}, we show the evolution of shell structure in Ca isotopes. At neutron number (N) =  20, the energy gap between $1p_{3/2}$ and $0f_{7/2}$ orbits is 2.72 MeV, which increases to 4.73 MeV as neutrons occupy $f_{7/2}$ orbit. This large gap is a standard picture of N = 28 magic shell gap that is observed in beta-stable and their nearby nuclei. In exotic Ca isotopes, the energy gap between $1p_{1/2}$ and $1p_{3/2}$ orbits enhances as neutrons occupy $1p_{3/2}$ orbit. It engenders semimagic character to $^{52}$Ca \cite{Gade,Wienholtz}. The separation between $0f_{5/2}$ and $0p_{1/2}$ orbits notably increases at N = 34. It supports the presence of semimagic shell gap for $^{54}$Ca \cite{Steppenbeck,Michimasa}.

\begin{table}
\caption{\label{tab:table1} Contribution of central, spin-orbit and tensor force components to the neutron single-particle energy gaps $1p_{3/2}-0f_{7/2}$,  $1p_{1/2}-1p_{3/2}$, and $1p_{1/2}-0f_{5/2}$ at N = 28, 32, and 34, respectively. All numerical values are given in MeV.}
\centering
\begin{ruledtabular}
	\label{T1}
	\begin{tabular}{lccc}
	 & $ 1p_{3/2}- 0f_{7/2} $ & $ 1p_{1/2}- 1p_{3/2}$ & $ 0f_{5/2} -  1p_{1/2} $\\
		& (N = 28) & (N = 32) & (N = 34) \\
		\cline{2-4}
		Central    &    0.842    &    0.255  &    0.640  \\
		Spin-orbit    &    1.192   &    0.374  &    -0.382 \\
		Tensor    &    -0.031    &    -0.179  &    -0.106     \\
		Total    &    2.004   &    0.449    &    0.152  \\
	\end{tabular}
\end{ruledtabular}
\end{table}

\begin{figure*}
\centering
\includegraphics [height= 7.5cm, width = 20cm, trim = {1.5cm, 5cm, 11cm, 1.cm}]{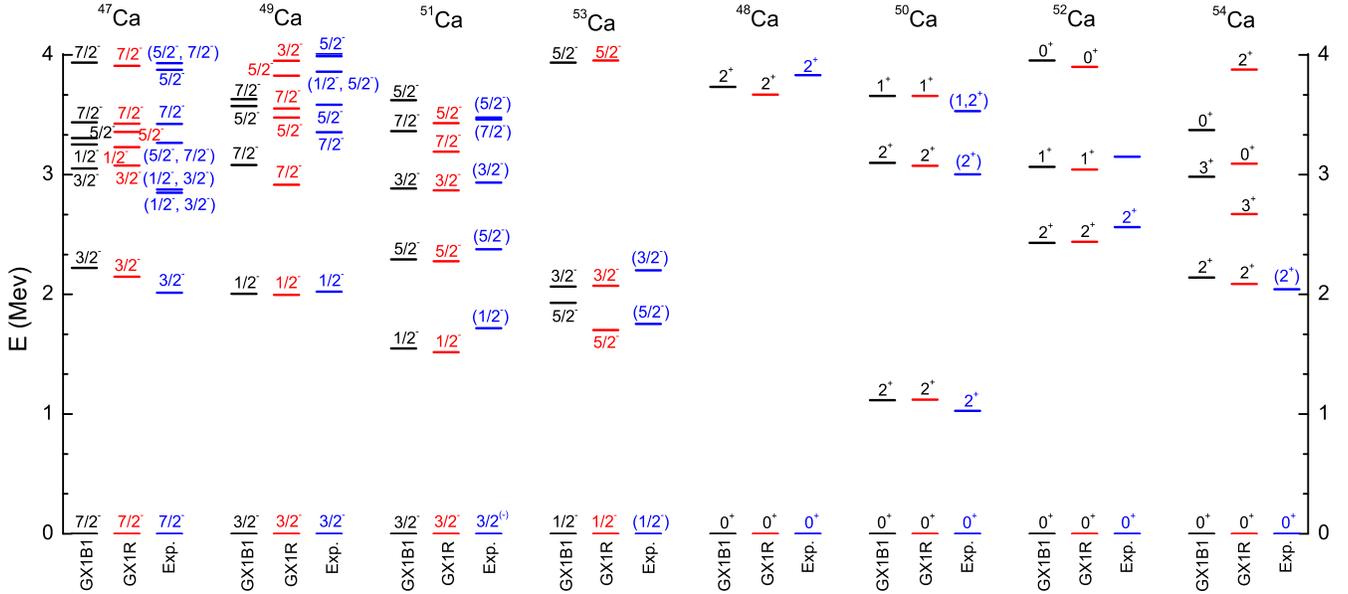}
\caption {Level structure of $^{47-54}$Ca.}
\label{F4}
\end{figure*}

In Table~\ref{T1}, we summarize the contribution of different components of GX1R to above-discussed single-particle energy gaps. The central and spin-orbit force, with nearly equal contribution, increase $1p_{3/2}-0f_{7/2}$ gap at N = 28. The $1p_{1/2}-1p_{3/2}$ gap at N = 32  has similar effects of  central and spin-orbit force. The contribution of tensor force in the same energy gap is relatively non-negligible, which tries to narrow it. In the development of $1f_{5/2}-1p_{1/2}$ gap as a semimagic gap at N = 34, the central force plays a crucial role. In the same gap, spin-orbit force acts opposite to central force with nearly half-strength.

The ordering of single-particle orbits in Ca isotopes is also interesting. There, $1f_{5/2}$ lies above $1p_{1/2}$ \cite{Uozumi}, which is not a case of the spherical mean-filed determined ordering of single-particle orbits \cite{Mayer,Bohr}. As an example, we study the evolution of ESPE of these orbits from Ca to Ni isotopes for N = 34. Results are shown in Fig.~\ref{F3}. It exhibits that the ordering gets normal for Ni. However, N = 34 semimagic shell gap disappears. For this change, central force is found to have dominant contribution.

\subsection{level structure}

The level structure of \textit{pf-}shell nuclei is very rich, and in the present case turns out as one mean to test the prediction power of GX1R interaction. Theoretical calculations have been performed for Ca, Ti, Cr, Fe, Ni, Co, Zn, and Ge isotopes with shell-model code NUSHELLX@MSU \cite{Brown4}. Results up to Cr isotopes are obtained without any truncation. For Fe, Ni, and Co, four particles are restricted in proton and neutron $0f_{7/2}$ orbit, whereas, for Zn and Ge isotopes, five particles are restricted. Experimental data is taken from \cite{nndc}.

\subsubsection{Ca isotopes}

In Fig~\ref{F4}, we show the level structure for $^{47-54}$Ca isotopes. Overall, comparison between theory and experiment is found to be good. For instance, in $^{47}$Ca, $\frac{3}{2}_{1}^{-}$ state measured at 2.01 MeV is predicted at 2.14 MeV.  In $^{49,51}$Ca,  $\frac{1}{2}_{1}^{-}$ state is found close to experimental data. This state is dominated by $\nu p_{1/2}^{1} \otimes \nu f_{7/2} ^{8}$ and $\nu p_{1/2}^{1} \otimes \nu p_{3/2}^{2}\nu f_{7/2} ^{8}$ configurations with 89.84\% and 85.33\% contribution, respectively. The high excitation energy of this state is linked to the N = 32 semimagic shell gap \cite{Rejmund}. In $^{53}$Ca, the $\frac{5}{2}_{1}^{-}$ and $\frac{3}{2}_{1}^{-}$ states are nicely reproduced. Here, $\frac{5}{2}_{1}^{-}$ state is dominated by $\nu f_{5/2}^{1} \otimes \nu p_{3/2}^{4}\nu f_{7/2} ^{8}$ configuration with 85.31\% contribution, and owns the signature of N = 34 semimagic shell gap \cite{Chen}. In case of even-even isotopes that consist of magic, semimagic, and open-shell nuclei, the measured E$(2^{+})$ is well predicted. In the experimental data, spin-parity of  third excited state of $^{52}$Ca is not known. In the calculation, it comes as 1$^{+}$.

\subsubsection{ Evolution of E$(2^{+}_{1})$}
\begin{figure}
\centering
\includegraphics [height =11cm, width=\columnwidth, trim={1cm 8.5cm 6.5cm 2cm}]{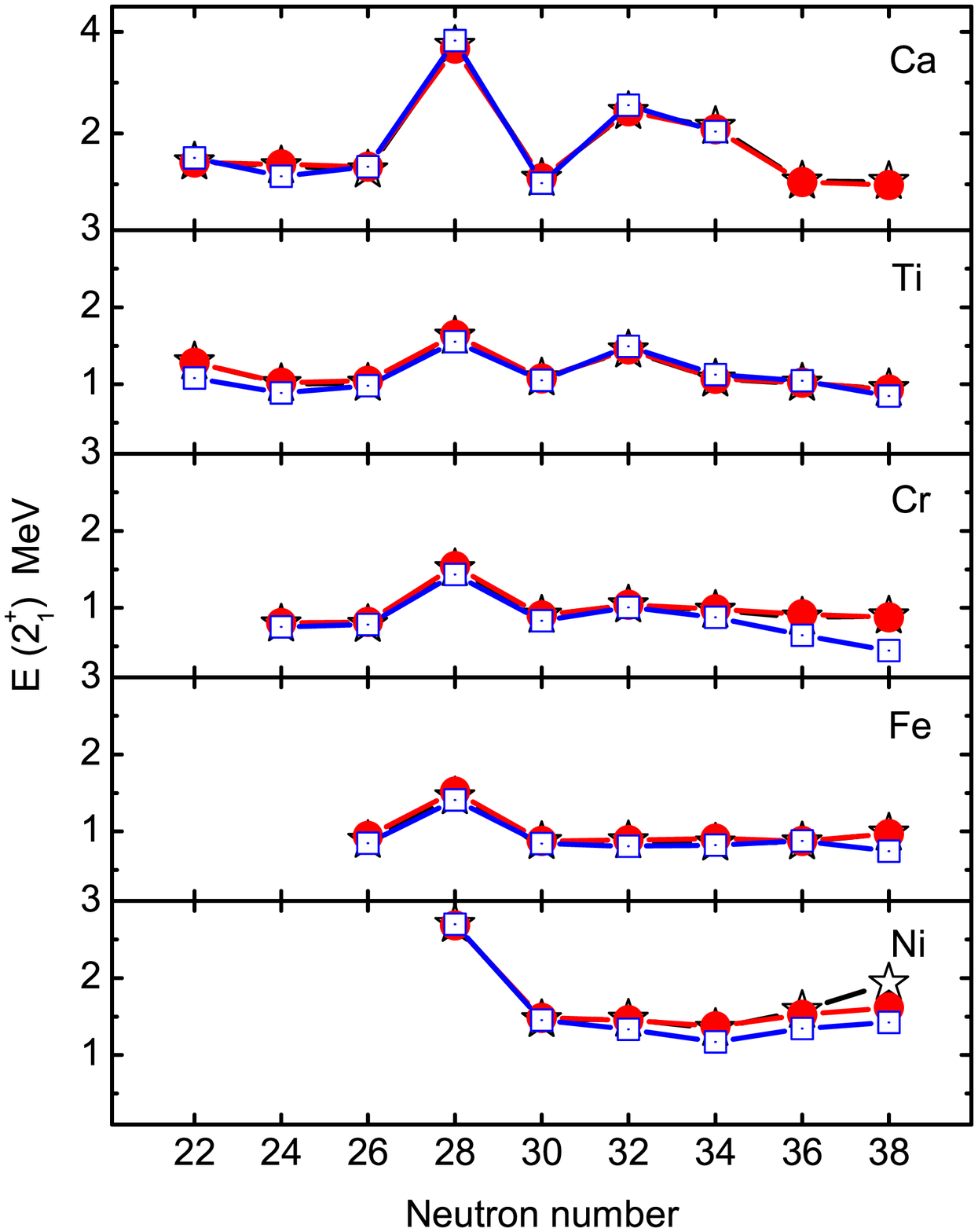}
\caption {Systematic of E$(2^{+}_{1})$ for Ca, Ti, Cr, Fe, and Ni isotopes. Results of GX1R and GX1B1 are shown by solid circle and open star, respectively. Experimental data is shown by open-square.  Only mirror asymmetry nuclei are considered.  }
\label{F5}
\end{figure}

The salient closed-shell features can be studied through the systematic of E$(2^{+}_{1})$ of even-even nuclei. The evolution of E$(2^{+}_{1})$ in Ca, Ti, Cr, Fe, and Ni isotopes is shown in Fig.~\ref{F5}.  The first kink in all isotope chains are caused by the N = 28 magic shell gap. The second kink in Ca isotopes possess the signature of N = 32 semimagic shell gap \cite{Gade}. The height of this kink reduces for heavier nuclei which indicates that N = 32 gap decreases with increase of Z, and consequently the quadrupole collectivity enhances. The high E$(2^{+}_{1})$ of $^{54}$Ca exhibits N = 34 semimagic shell gap effect \cite{Steppenbeck}. This E$(2^{+}_{1})$ reduces to half of its value for $^{56}$Ti, which is nearly equal to E($2^{+}_{1}$) of open-shell isotopes $^{52,58}$Ti. Thus, it shows that N = 34 energy gap has fragile nature, and melts down for $^{56}$Ti. In Fig.~\ref{F3}, we have predicted the similar evolution of N = 34 gap for $Z>20$ nuclei. In theoretical study \cite{Utsuno3}, however, it has been reported that N = 34 gap gets strong for $Z<20$ nuclei. Recent experimental results of $^{52}$Ar also support this fact \cite{Liu}.

Theoretical results are found to be in good agreement with experimental data in Fig.~\ref{F5}, except for a few N$\sim $40 nuclei. For $^{60-64}$Cr, $^{64}$Fe, and $^{64, 66}$Ni nuclei, the difference is found in the range of 0.2-0.4 MeV. This indicates that the orbit(s) from higher shell, such as $\nu 0g_{9/2}$, is(are) required to describe the properties of heavy $\pi 0f_{7/2}$ nuclei \cite{Lenzi}.

\subsubsection{Softness of $^{56}$Ni core}

\begin{figure}[h!]
\centering
\includegraphics [height= 8.1cm, width = \columnwidth, trim = {2.5cm, 1.5cm, 7.8cm, 0cm}]{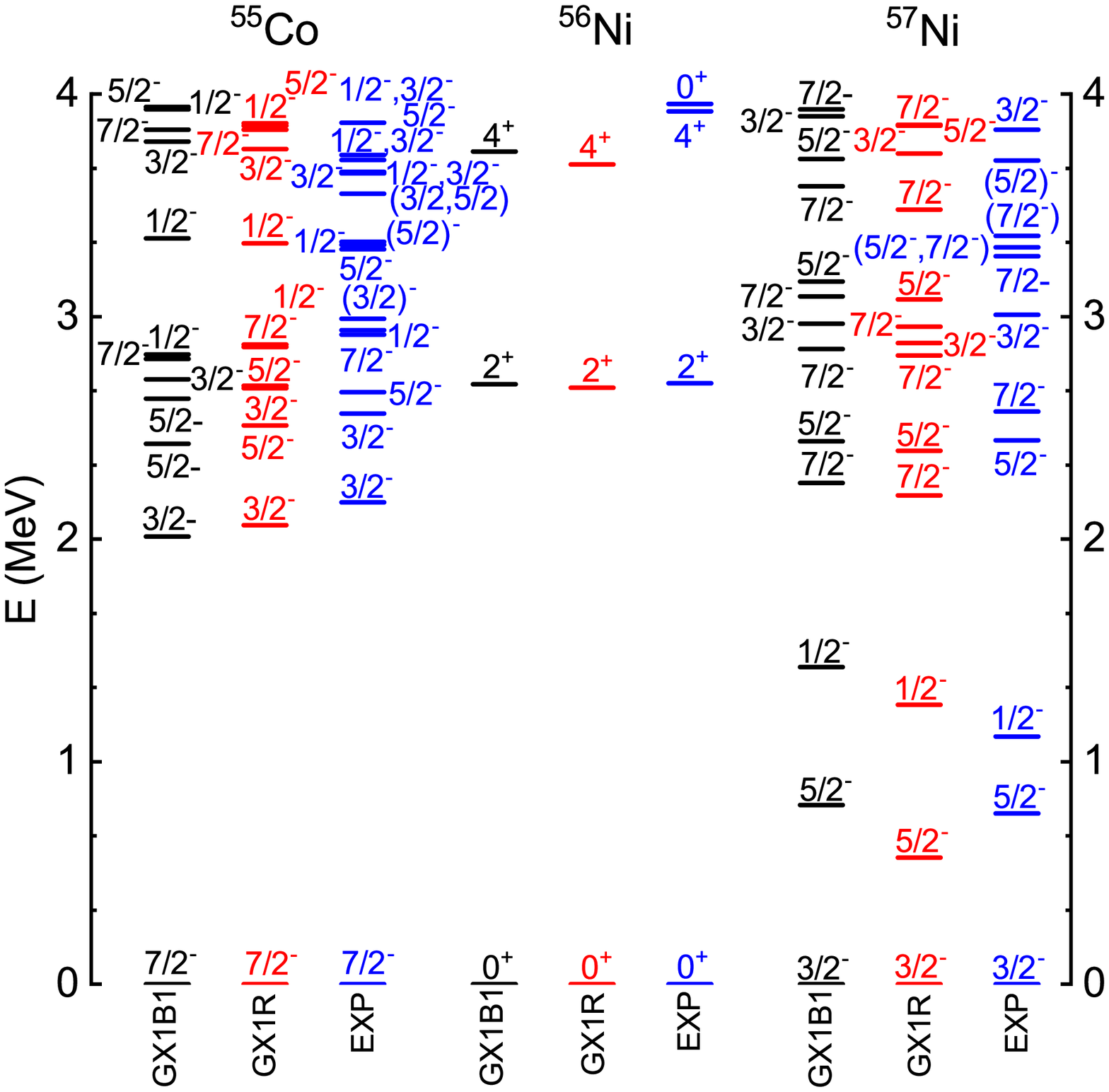}
\caption {Level structure of $^{55}$Co, and $^{56,57}$Ni.}
\label{F6}
\end{figure}

The softness of $^{56}$Ni core has been one of the challenges for $pf-$shell interactions due to its substantial effects on the level structure of $^{56}$Ni and its nearby nuclei\cite{Poves2,Honma3,Ritcher}. The old interactions were insufficient to describe the properties of these nuclei, which eventually led to the construction of GXPF1 \cite{Honma3}. In this section, we study the properties of low-lying states of $^{55}$Co and $^{56,57}$Ni, and estimate the size of core-component in them. The level structure of these nuclei is shown in Fig.~\ref{F6}. The agreement between theory and experiment is found to be fair.

The ground state and the first $\frac{3}{2}^-$ and $\frac{5}{2}^-$ states of $^{55}$Co can be described as the one-hole (1h) and the one-particle two-hole (1p-2h) proton-configuration with respect to $^{56}$Ni core, respectively. In the results, the corresponding proton-configuration has  60\% contribution for ground state and about 27\% for  first $\frac{3}{2}^-$ and $\frac{5}{2}^-$ states. In the ground state of $^{56}$Ni, the core is strong as 67\%. This number is relatively smaller than $^{48}$Ca core (93\%), which manifests the softness of doubly magic $^{56}$Ni core. The $2^{+}_{1}$ and $4^{+}_{1}$ states have 1p-1h character with 41-44\% broken-core component. We have noticed in the calculations that if the core is kept strong, the $2^{+}_{1}$ moves up in energy and deteriorates agreement with the experiment. The $^{57}$Ni has one-particle configuration with respect to $^{56}$Ni in the single-particle shell model. Thus, the ground state $\frac{3}{2}^-$, and the first $\frac{5}{2}^-$ and $\frac{1}{2}^-$ states can be ascribed as the single-particle states, corresponding to neutron in $1p_{3/2}$, $0f_{5/2}$, and $1p_{/2}$ orbit, respectively. In full shell-model results, these states have 53\%, 51\%, and 36\% single-particle strength, respectively. For other excited states, the single-particle strength decreases further. 

\subsubsection{$Z>28$ nuclei}

The Z$>$28 nuclei are usually studied in the reference of $f_{5/2}pg_{9/2}-$model space \cite{Honma4,Brown5} , however, the data of these nuclei is included for driving $pf$-shell interaction \cite{Honma3}. Therefore, their results also need to be investigated in $pf$-shell to account the righteousness of an interaction on a broader scale. In Fig.~\ref{F8}, we present the systematic of E$(2^{+}_{1})$ for Zn and Ge isotopes. As can be seen, GX1R well produces constant E$(2^{+}_{1})$ for both sets of nuclei.

\begin{figure} [h!]
\centering
\includegraphics [height= 5.2cm, width = \columnwidth, trim = {0.7cm, 4.1cm, 7.1cm, 0.8cm}]{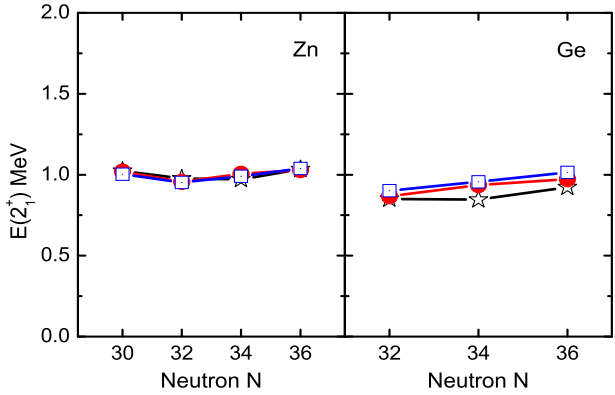}
\caption {Systematic of E$(2^+_1)$ for Zn and Ge isotopes. See caption of Fig.~\ref{F5} for symbols.}
\label{F7}
\end{figure}


\subsection{GX1R vs. GX1B1}

\begin{figure}
\centering
\includegraphics [height= 4.9cm, width = \columnwidth, trim = {2cm, 3.5cm, 8.3cm, 0.8cm}]{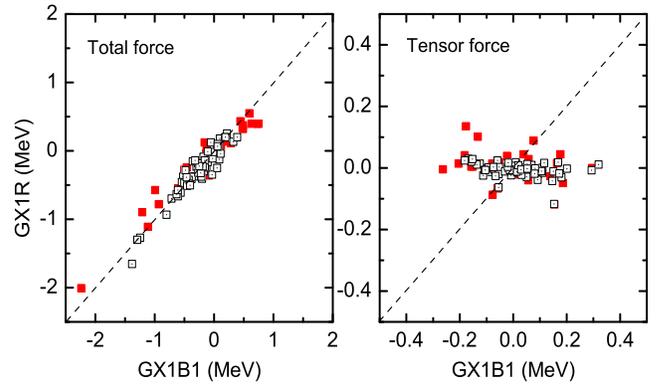}
\caption {Comparison of matrix elements of GX1R and GX1B1 interactions. It is shown for total force on the left, and for tensor force on the right. Diagonal and non-diagonal matrix elements are shown by solid and open square, respectively.}
\label{F8}
\end{figure}

In Figs~\ref{F4}$-$\ref{F7}, we show results of GX1B1 interaction as well. They are obtained in same conditions, and found to be almost same as those of GX1R. In order to find the reason for this, we have compared the matrix elements of both the interactions. This comparison is presented in Fig.~\ref{F8}. Matrix elements mostly fall close to the diagonal-line which indicates that they are more or less similar in both interactions. Matrix elements of GX1R have 0.14 MeV root mean square deviation with respect to GX1B1. The comparison of tensor force matrix elements is also shown in Fig~\ref{F8}. Here, matrix elements deviate from diagonal-line. However, their magnitude is small. It can be one plausible reason for having similar matrix elements in GX1R and GX1B1.

In the context of present work, it is important to mention that there is no ambiguity on the predictive power of GX-interactions. These interactions are widely accepted, and fairly describe the properties of $pf-$shell nuclei. In this work, our prime aim is to correct the tensor force disparity of GX-interactions. From the comparison of results of GX1R and GX1B1, it can be deduced that GX1R is also an adequate effective interaction. This interaction has been recently used to calculate the nuclear matrix elements (NMEs) of neutrinoless double beta decay process for $^{48}$Ca \cite{Sarkar}. The results shows only $1-3$\% increase in NMEs with respect to GXPF1A.

\section{Summary}
\label{S4}

In this work, we have examined the systematic properties of tensor force in  GX-interactions and discerned that seven out of ten T = 1 monopole matrix elements do not possess those properties. We have corrected this irregularity by replacing all ninety-four tensor force TBMEs of GX1B1 interaction with the semi-empirically calculated ones using spin-tensor decomposition. In an interesting way, calculated all ten T = 0 and three T = 1 $1p-$orbit tensor force monopole matrix elements are found similar to those of GX1B1. This exhibits that the employed method is one of the ways that incorporates universal features of two-nucleon force, and missing in-medium and three-body force effects reflected in tensor component of the interactions.



With the additional modification of single-particle energy of $1p_{3/2}$ orbit and two TBMEs of $0f$-orbit, the derived interaction, as named GX1R, is employed to discuss the shell evolution in Ca isotopes and N = 34 isotones, level structure of Ca isotopes, evolution of E$(2^{+}_{1})$ in Ca, Ti, Cr, Fe and Ni chain, softness of $^{56}$Ni core, and E$(2^{+}_{1})$ systematic for Z$>$28 nuclei.  Results are found to be satisfactory with respect to experimental data. Calculations are also performed with GX1B1, and the results are observed to be similar to those of GX1R. Despite the difference of tensor force matrix elements, the total matrix elements of both interactions are found to be similar, making their predictive power almost similar.

\section{ACKNOWLEDGMENTS}
K. Jha thanks R. G. Pillay and P. K. Rath for their interest and useful scientific discussions. P. Kumar thanks P. P. Singh and S. Thakur for helpful suggestions in writing the manuscript.

\end{document}